\documentclass[aps,prl,twocolumn,groupedaddress,showpacs]{revtex4}

\usepackage{amsmath}
\usepackage{microtype}
\usepackage{subfigure}
\usepackage{amsfonts}
\usepackage{amssymb}
\usepackage{changepage}
\usepackage{graphicx}

\newcommand{\be}{\begin{equation}}
\newcommand{\ee}{\end{equation}}

\newcommand{\rd}{{\rm d}}

\newcommand{\cH}{{\cal H}}
\newcommand{\cL}{{\cal L}}

\newcommand{\Ym}{{\sf Y}}
\newcommand{\sP}{{\sf P}}
\newcommand{\Sx}{{\sf S}_{\times}}

\newcommand{\lc}{{\ell_\mathrm{c}}}

\def\lsim{\mathrel{\rlap{\lower4pt\hbox{$\sim$}}
    \raise1pt\hbox{$<$}}}                % less than or approx. symbol
\def\gsim{\mathrel{\rlap{\lower4pt\hbox{$\sim$}}
    \raise1pt\hbox{$>$}}}           

\begin{document}
\preprint{APS/123-QED}

\title{Curvature-induced cross-hatched order\\ in two-dimensional semiflexible polymer networks}% Force line breaks with \\

\author{Cyril Vrusch}
%\affiliation{
% Department of Applied Physics and Institute for Complex Molecular Systems, Eindhoven University of Technology,\\
% P.O. Box 513, NL-5600 MB Eindhoven, The Netherlands
%}%
\author{Cornelis Storm}
\affiliation{
 Department of Applied Physics and Institute for Complex Molecular Systems, Eindhoven University of Technology,\\
 P.O. Box 513, NL-5600 MB Eindhoven, The Netherlands
}%


%\affiliation{
%\affiliation{
%\affiliation{%


\date{\today}% It is always \today, today,

\begin{abstract}
A recurring motif in the organization of biological tissues are networks of long, fibrillar protein strands effectively confined to cylindrical surfaces. Often, the fibers in such curved, quasi-2D geometries adopt a characteristic order: the fibers wrap around the central axis at an angle which varies with radius and, in several cases, is strongly bimodally distributed. In this Letter, we investigate the general problem of a 2D crosslinked network of semiflexible fibers confined to a cylindrical substrate, and demonstrate that in such systems the trade-off between bending and stretching energies, very generically, gives rise to cross-hatched order. We discuss its general dependency on the radius of the confining cylinder, and present an intuitive model that illustrates the basic physical principle of curvature-induced order. Our findings shed new light on the potential origin of some curiously universal fiber orientational distributions in tissue biology, and suggests novel ways in which synthetic polymeric soft materials may be instructed or programmed to exhibit preselected macromolecular ordering.

%\begin{description}
%\end{description}
\end{abstract}

\maketitle
\begin{figure}[t!]
\begin{center} 
\includegraphics[width=\columnwidth,height=2.1cm]{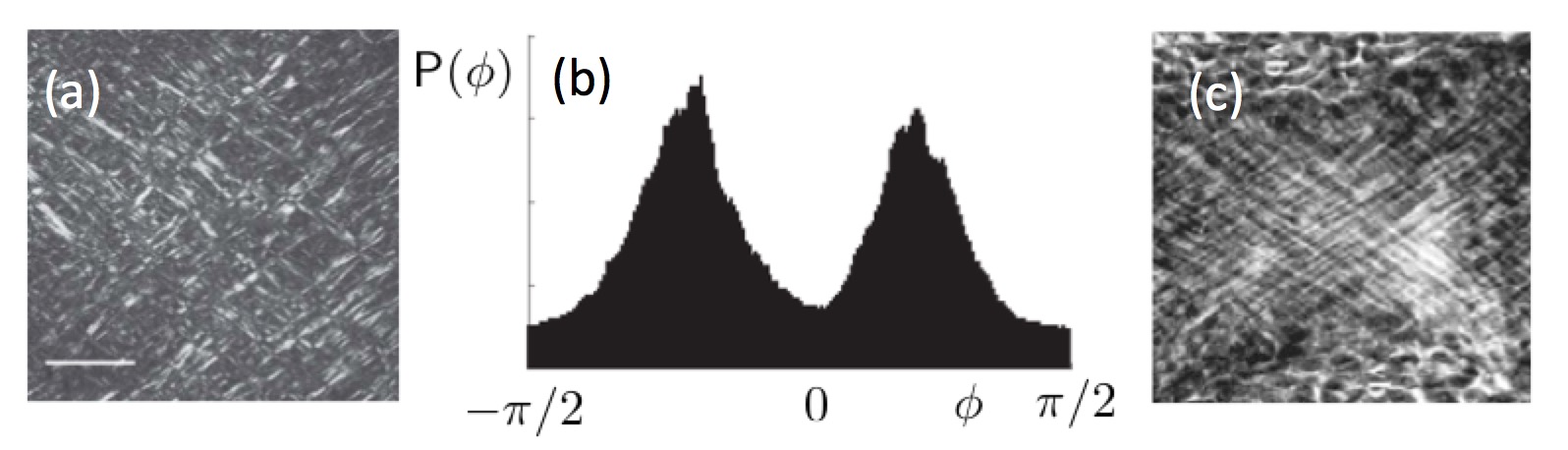}
\end{center}
\caption{(a) Cross-hatched order of collagen in the intima of the human thoracic aorta \cite{schriefl2012}. (b) The orientational distribution function for the angle $\phi$ with the cylindrical axis of the fibers in image (a). (c) Cross-hatched ordering of collagen in the midsagittal section of the annulus fibrosus \cite{hayes1999}.}
\label{Fig1}
\end{figure}

In the arterial wall, quasi-2D layers of meshed collagen fibers alternate with layers of smooth muscle cells to envelop and strengthen the cylindrical lumen of the blood vessel \cite{schriefl2012}. Likewise, in the annulus fibrosus of the intervertebral disc concentric, thin lamellae of collagen cylindrically surround the soft nucleus pulposus \cite{marchand1990,benneker2004,hayes1999}, an ordering that is proposed to find its origin in the collagen sheet that wraps around the notochord, the embryonal precursor of the spine \cite{adams1990mechanics}. Elsewhere in the body, in cortical bone thin lamellae of mineralized collagen are wrapped cylindrically around each Haversian canal in a structure called the osteon \cite{giraud1988,wagermaier2006spiral,carnelli2013orientation}. In each of these, physiologically unrelated, settings a universal organization is employed: thin networks of long, fibrillar protein strands tightly enveloping cylindrical domains. Remarkaby, in each of these cases, fibers in such curved, quasi-2D geometries tend to become ordered: the biopolymers wrap around the central axis at an angle which varies with radius and, in several cases, is strongly bimodally distributed, leading to a cross-hatched appearance. Several examples of such order are depicted in Fig. \ref{Fig1}.

Order in filamentous biomaterials may be provoked in several manners. In \cite{vader2009}, it is reported strain may induce alignment. Part of this alignment is imprinted irreversibly. In \cite{flynn2010,bhole2009}, it is demonstrated that strain, likewise, may affect the enzymatic degredation of collagen in living tissues. Strain-dependent fiber deposition and degradation - mimicking cellular or enzymatic activity - is included in several constitutive models \cite{kuhl2007,driessen2004,driessen2008}, and the interactions between cylinders (carbon nanotubes and histones, in particular) and polymers (either single polymers, or dilute solutions without linking) have been adressed in several previous works, reporting helical wrapping configurations - but in these cases the polymers were not fully confined to the surface \cite{kusner2006,gurevitch2007,vazquez2015} but rather interacted with it through a weak attractive potential. Though these models all display some degree of ordering, they fail to provide a mechanism for this ordering in the absence of active orienting entities such as cells - the manner in which it appears to occur in many biological systems. We do not claim that in all of the biological instances we have mentioned a single, physical mechanism explains all of the observed structure (indeed, in several cases the cross-hatched order is absent, or replaced by more helical patterns).  What is, however, clear is that the interaction between curved surfaces and semiflexible polymers, in general, gives rise to nontrivial ordering effects that are manifested in several, biologically relevant, settings. 

In order to understand and, perhaps, learn how to control this ordering we investigate a simple model of a polymer meshwork fully confined to the surface of a cylinder.

{\it Model}. Our polymer chains are confined to the surface of a cylinder. We shall denote by $(u,v)$ the two coordinates on the rolled-out cylinder with $0<u<L$ the axial coordinate, and $0<v<2\pi$ the circumferential coordinate; these coordinates are mapped in 3D onto the surface of a cylinder of radius $R$ and length $L$ lying along the $y$-axis: $\vec r(u,v)\equiv\{x(u,v),y(u,v),z(u,v)\}=\{R \cos(v/R), u, R \sin(v/R)\}$. The semiflexible polymers themselves are modeled as discrete, extensible persistent chains. That is, chains of total, unstretched contour length $\lc$ are represented by discrete segments labeled by $i=1,\cdots,N$, each such segment having a different rest length $\ell_{0,i}$. Segment $i$ follows bond 2-vector $\vec b_i=\{u_{i+1}-u_i,v_{i+1}-v_i\}$ on the rolled-out cylinder, and bond 3-vector $\vec \beta_i=\vec r(u_{i+1},v_{i+1})-\vec r(u_{i},v_{i})$ in 3D. The 2- and 3-lengths of segments are denoted by $\ell_i=(\vec b_i \cdot \vec b_i)^{1/2}$ and $\lambda_i=(\vec \beta_i \cdot \vec \beta_i)^{1/2}$, respectively. All segments possess the same extensional modulus $\Ym$, and correspondingly have different spring constants $k_i=\Ym/\ell_{0,i}$. Stretching of each segment is modeled as a simple Hookean spring, with a stretching energy $\cH_{{\rm stretch},i}=\frac{\Ym}{2 \ell_{0,i}} (\ell_i-\ell_{0,i})^2$. For the bending energy of the chain, there are two contributions that one must consider. The first is the bending between segments, proportional to the square of the angle between two adjacent segments. The second, however, arises uniquely on curved substrates and measures the bending of each segment individually. First, a note on the intersegment bending term. This term, proportional to the cosine of the angle between subsequent segments, may be written in terms of the angle $\Xi_i$ between adjacent bond vectors $\vec \beta_i$ and $\vec \beta_{i+1}$ as $\cH_{{\rm interseg},i}= \frac{2 \kappa}{\lambda_i+\lambda_{i+1}} \left(1-\cos\Xi_i\right)$. However, we will consider here networks {\em fully confined} to the cylindrical surface, and will not allow for excursions or shortcuts through the interior. That is, we will assume that $\ell_i=\lambda_i$, as the mapping onto the cylinder surface is isometric. A consequence of this is that each individual segment is bent to conform to the surface - the additional energy this imparts is treated later on. We may now write the angle bending term, instead, as a function of the angle $\theta_i$ between segments in the $(u,v)$ plane. Due to the isometric mapping the in-plane angle $\theta$ at which two bent segments join remains the same as the 3D angle $\Xi$ (this is no longer true for general curved surfaces). Thus, our angle bending potential is chosen as
\be
\cH_{{\rm interseg},\{i,i+1\}}=\frac{2 \kappa}{\ell_i+\ell_{i+1}} \left(1-\cos\theta_i\right)\approx \frac{\kappa}{\ell_i+\ell_{i+1}}\theta_i^2\, ,
\ee
where in the final equation we assume small in-plane angles between the fibers. As stated, the energy of the chain picks up an additional segment contribution, which measures the bending of segments to conform to the surface. This we compute by first constructing the geodesic $\vec \gamma_i(s)$, connecting the points $(u_i,v_i)$ and $(u_{i+1},v_{i+1})$, with $0<s<\ell_i$ its natural (arclength) parameter. The geodesic is a simple linear form in $s$ in this case, and the curvature contribution is then straighforwardly (and analytically) evaluated to be
\be
\cH_{{\rm segbend},i}=\frac{\kappa}{2}\int_0^{\ell_i} \left|\frac{\rd^2 \vec \gamma_i}{\rd s^2}\right|^2\rd s=\frac{\kappa}{2} \frac{\ell_i}{R^2} \sin^4 (\phi_i)\, ,
\ee
with $\phi_i$ the angle between the segment and the $y$-axis (i.e., the axial direction of the cylindrical substrate): $\cos(\phi_i)=\hat b_i\!\cdot\!\hat u$.

The appearance of this angle should come as no surprise - a segment aligned to the long axis of the cylinder ($\phi=0$) remains unbent, whereas one that is circumferentially oriented ($\phi=\pm \pi/2)$ is maximally bent, when wrapped around the cylinder. This is the origin of the effects we discuss here: the curved substrate acts, locally, as an external field in the $\pm \hat u$-directions. Obviously, the term vansishes as $R^{-2}$ when the radius of the cylinder is increased. While one might be tempted to expand the expression for $\cH_{\rm segbend}$ around $\phi=0$, this is not justified as the random arrangement of chains in a crosslinked network consists of segments oriented at all angles with respect to the cylinder's long axis. Thus, our complete energy functional is
\begin{eqnarray}
\cH&=&\sum_{i=1}^{N}\left(\frac{\Ym}{2 \ell_{0,i}} (\ell_i-\ell_{0,i})^2+\frac{\kappa}{2} \frac{\ell_i}{R^2} \sin^4 (\phi_i) \right)+\nonumber \\ & &\qquad +\sum_{i=1}^{N-1}\left(\frac{\kappa}{\ell_i+\ell_{i+1}}\theta_i^2\right)\, .
\label{toten}\end{eqnarray}
In what follows, we attempt to characterize the typical spatial arrangement of the fibers in networks that minimize this energy functional. Before we move on to crosslinked networks, it is instructive to first consider the highly simplified model of a single ring polymer wrapping around a cylinder.

\begin{figure}[t!]
\begin{center} 
\includegraphics[width=\columnwidth,height=2.25cm]{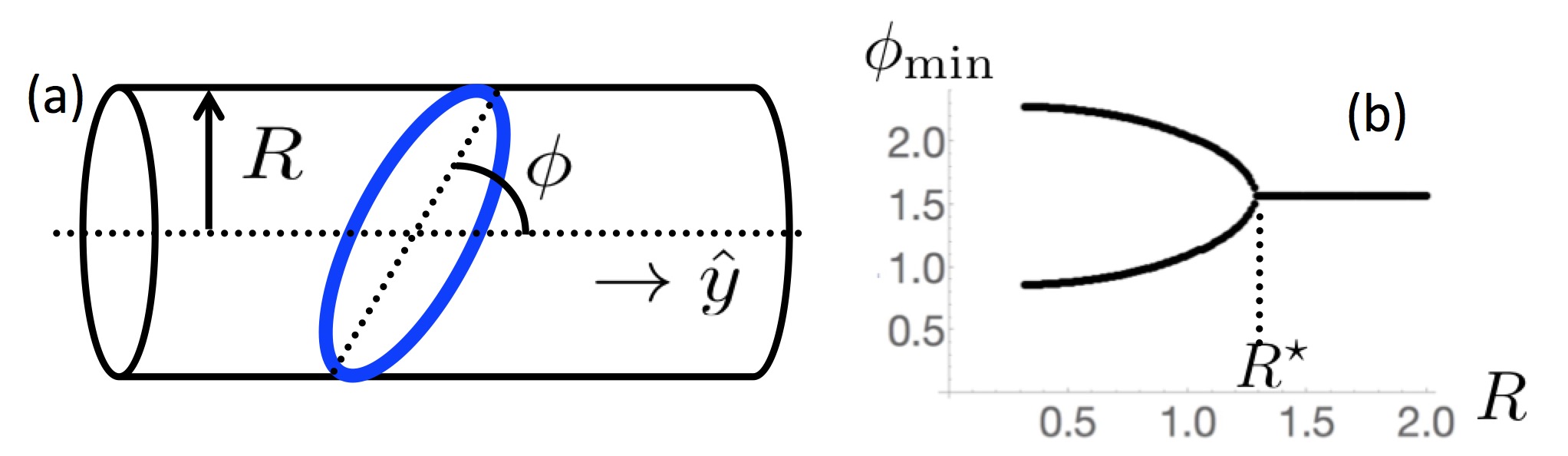}
\end{center}
\caption{(a) The ring model: a semiflexible ring polymer (blue) is wrapped around a cylinder of radius $R$, at an angle $\phi$ with respect to the cylinder's long axis. A competition between bending (favoring axial alignment) and stretching (favoring circumferential alignment) results in a bifurcation. (b) The optimal angle $\phi_{\rm min}$ as a function of the radius for a polymer of length $L_0=2$.}
\label{Fig2}
\end{figure}

{\it Ring Polymer.} For a simple, continuous ring polymer wrapped around a cylinder of radius $R$, the problem and the typical nature of solutions becomes clear. We will assume, again, that the polymer has a rest length $\ell_0$, a spring constant $\Ym/\ell_0$ and a bending modulus $\kappa$. The tradeoff for this ring polymer, in terms of energy contributions, is obvious: circumferential positioning allows the polymer minimal length, but at the expense of maximal curvature (since the radius $R$ is the only curvature length scale in the problem). Slanted arrangements, where the polymer traces out an ellipse that forms the intersection between the cylinder and a plane at some angle, require less curvature energy but more stretching. We call $\phi$ the angle of the ring polymer with respect to the cylinder's long axis (i.e., $\phi=\pm\pi/2$ corresponds to circumferential placement, and $\phi=0$ to infinitely stretched out axial placement), and call $C_\phi$ the corresponding elliptic conformation. Since the polymer remains planar, $C_\phi$ may be parametrically represented in 2D as $C_\phi(t)=R \{\sin^{-1}(\phi)\cos(t),\sin(t)\}$; $0<t<2 \pi$. We will denote by $\cL_\phi$ the arc length corresponding to one trip around the cylinder at angle $\phi$ (a closed form expression in terms of elliptic integrals exists for $\cL_\phi$, but this is not particularly instructive). The optimal angle $\phi$, given $\Ym$ and $\kappa$, may be determined now by minimizing over all $\phi$ the energy
\be
\cH_{\rm ring}(\phi)=\frac{\Ym}{2 \ell_0}(\cL_\phi-\ell_0)^2+\frac{\kappa}{2}\oint_{C_\phi}\left|\frac{\rd^2 C_\phi}{\rd s^2}\right|^2\rd s\, .
\ee
Evaluating this expression, and minimizing it as a function of $\phi$ yields an interesting result, depicted in Fig. \ref{Fig2}. For large radii of the cylinder, the bending terms become unimportant and only the stretching remains. 
The stretching terms simply seeks to minimize the contour length, and indeed the energy is minimized at $\phi=\pi/2$; the ring encircles the cylinder orthogonal to its long axis. If the radius is lowered, however, beyond a critical radius $R^\star$, the bending becomes increasingly important and the circumferential ordering gives way to two related slanted orientations as the new minima. The critical radius $R^\star$ may be expressed in dimensionless form: $R^\star/\ell_0=\psi(\Ym \ell_0^2/\kappa)$ with $\psi(x)$ a universal, though rather unwieldy, function of $x$. The upshot of this simple ring model is, that for closed filaments confined to the surface of a cylinder, below a critical radius the circumferential ordering is superseded as the minimal energy configuration by two slanted orientations, whose angles $\phi$ are symmetric around $\pi/2$. Clearly, this reminds us of the crossed order that is seen in the biological systems we recalled in the introduction. We now turn to simulations of fully crosslinked {\em networks} of polymers to verify to what extent this behavior is manifested in more complex settings. 

{\it Network simulations}. To simulate the behavior of many polymers, crosslinked to each other and confined to the surface of a cylinder, we adopt the mikado procedure: an initial network of straight fibers is prepared by randomly distributing them on the rolled-out cylinder (i.e., in the $(u,v)$-plane). The chains interact with each other only at crosslinking points, there is no excluded volume interaction. Dangling ends are removed. In this initial network, the energy contributions from stretching $\cH_{\rm stretch}$ and bending $\cH_{\rm interseg}$ are zero by construction. The segment bending term $\cH_{\rm segbend}$, however, is not. While any such mikado arrangement is a minimizer of the energy functional $\cH$ for $R=\infty$, this is not true for finite radii. Therefore, we allow the initial network to relax to a minimal energy state, in which some of the segment bending energy is alleviated by reorienting the segments. The elastic energy lost in doing so is, of course, partly traded for stretching and intersegment bending energies. To find the energy minimum, we employ a multistep non-linear conjugate gradient scheme \cite{shewchuk1994introduction}.

\begin{figure}[b]
\begin{center} 
\includegraphics[width=\columnwidth]{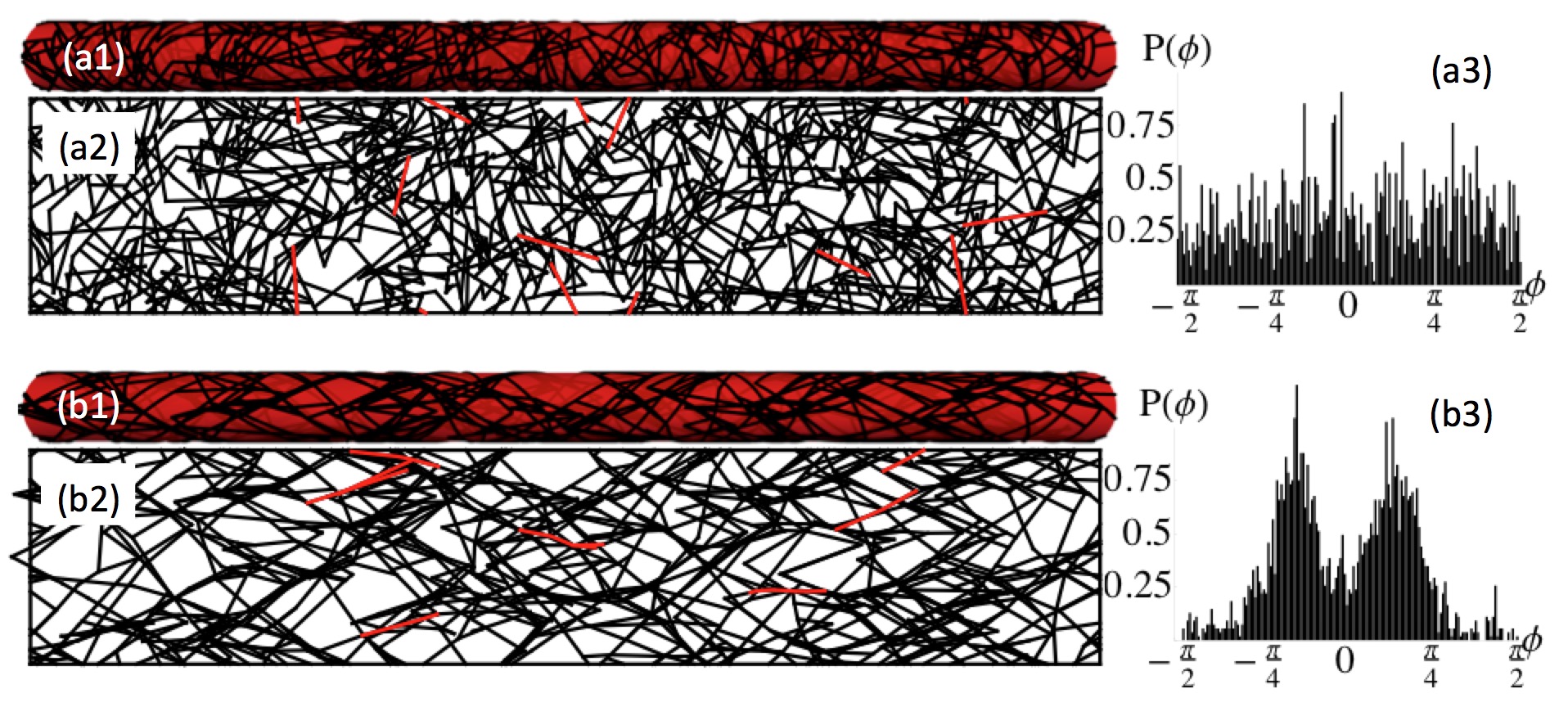}
\end{center}
\caption{(a1) The initial mikado network, wrapped around the cylinder. (a2) the same, rolled out onto the $(u,v)$-plane. (a3) the initial angle distribution $\sP(\phi)$ is flat. (b1,2,3): same images, for the network configuration that minimizes Eq. \ref{toten}}
\label{Fig3}
\end{figure}

The result of such a minimization is shown in Fig. \ref{Fig3}. Figs 3(a1), 3(a2) and 3(a3) display data from the initial, disordered configuration, where $\cH_{\rm segbend}$ is the only non-zero energy component. Figs. 3(b1), 3(b2) and 3(b3) present the network as it appears after the minimization procedure. The network shown here consists, before pruning, of 800 fibers with $\ell_0=0.9$, the radius and length of the cylinder are $R=\pi^{-1}$ and $L=10$. The filaments are fixed at the ends of the cylinder. The effective spring constant $\Ym=1$, and $\kappa=10$. One may glean from Figs 3(a1),(a2),(b1) and (b2) (which show the network on the cylinder, and rolled out onto the $(u,v)$-plane) in particular, there is considerable reorientation in the relaxed network, which has an overall energy that that is a factor 2.5 lower than the initial network. Several filaments have been plotted in red, to illustrate that initially straight filaments are now bent. Also, it is apparent from the increased size of the open spaces that - while the overall density has not changed - there is considerable stretch and compression present in the relaxed network. The most striking change, however, is manifested in the orientational distribution $\sP(\phi)$, which gives the probability of a segment's angle with respect to the long axis being equal to $\phi$, normalized over the interval $-\pi/2<\phi<\pi/2$. Where initially, this distribution is uniformly flat ($\sP(\phi)=\pi^{-1}$, by definition for an unstrained mikado network), it acquires a marked bimodal structure, with two peaks symmetrically located around $\phi=0$, see Figs 3(a3) and 3(b3). For lack - to our knowledge - of an existing term we shall dub this type of order {\em cross-hatched}. It is distinct from the more common tetratic order, reported for several liquid crystalline systems \cite{jabbarzadeh2007,donev2006}. Tetratic phases, mostly seen in lyotropic systems, are composed of two distinct types of nematic domains with local directors at right angles to each other. A true tetratic phase has fourfold rotational symmetry. Though the tetratic phase represents a specific instance of cross-hatched order, it possesses a higher symmetry than the general cross-hatched phase, which has two {\em superposed} nematic phases, with two directors that are generally {\em not} oriented at right angles to each other and thus has two separate twofold rotational symmetries. 

To quantify the extent of the cross-hatched order (denoted CH, in the following), we introduce an order parameter that respects the symmetries of the CH state. First, we define the CH-director as the unit vector $\hat n={\vec\Delta}/|\vec \Delta|=(n_u,n_v)$, with

\begin{equation}
\vec \Delta =\int_{-\pi/2}^{\pi/2}\!\!\! \left(\mathbf{\hat{u}}\cos|\phi|+\mathbf{\hat{v}}\sin|\phi|\right) \sP(\phi)\rd \phi\,.
\end{equation}
The CH director points along the direction of the positive-$\phi$ peak in $\sP(\phi)$ for sufficiently narrow distributions, and is oriented at an angle $\langle\phi\rangle = \cos^{-1}(n_u)$ with respect to the cylinder's long axis. We now define the CH-order parameter $\Sx$ to be
\be
\Sx =\int_{-\pi/2}^{\pi/2}\!\!\! \cos\left(4(|\phi|-\langle\phi\rangle)\right) \sP(\phi)\rd \phi.
\ee
This order parameter $\Sx$ is zero in an isotropic network, and 1 for any perfectly cross-hatched network - {\em i.e.}, a network where $\sP(\phi)=(1/2)[\delta(\phi-\phi_0)+\delta(\phi+\phi_0)]$, regardless of the value of $\phi_0$. Similar to its conventional nematic counterpart (which is zero for a CH phase), $\Sx$ quantifies the extent of the CH order, but does not contain information regarding its spatial orientation. Clearly, the initial network as depicted in Figs. 3(a1) and 3(a2) has $\Sx=0$, the relaxed network depicted in Figs 3(b1) and 3(b2) has $\Sx=0.50$. Thus, the curvature of the substrate is seen to evoke, in this case, significant cross-hatched order. 

\begin{figure}[t]
\begin{center}
\includegraphics[width=\linewidth]{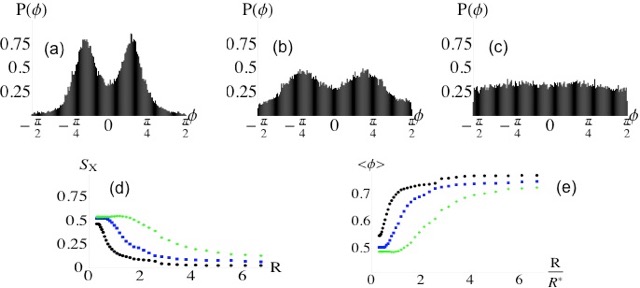}
\caption{Orientational distribution functions in energy-minimized networks on cylinders of radius (a) $R=1/\pi\equiv R_0$, (b) $R=6.1 R_0$, (c) $R=41.1 R_0$. The double peak is seen to dissappear at larger radii, as expected. In (d) and (e) the cross-hatched order parameter $S_X$ and $\langle\phi\rangle$, which is the orientation of the peak for highly ordered networks, is shown. The black, blue and green are the data from simulations with $1$, $10$ and $100$ for $\Ym$, respectively and $\kappa=1$.\label{Fig4}}
\end{center} 
\end{figure}

To determine the dependence of the curvature-induced CH order and orientations on the radius of the cylinder, we perform a series of simulations of the network on cylinders of increasing radius. The results are collected, for a representative choice of network parameters, and averaged over an ensemble of $20$ initial networks, in Fig. \ref{Fig4}. The first striking observation is the emergence of strongly bimodal orientational distribution at small cylinder radii. This is easily seen in $\sP(\phi)$, and likewise reflected in the order parameter. For large radii, we find disordered networks with $\Sx=0$, and $\langle\phi\rangle=\pi/4$. For smaller radii, we see $\Sx$ come up, as well as the emergence of average angles $\langle\phi\rangle$ smaller than $\pi/4$ (note, that a given value of $\langle\phi\rangle$ corresponds to two peaks in $\sP$, on either side of $\phi=0$). We see that the network is longer in an ordered state as $\Ym$ increases.

{\it Conclusions}. In this Letter, we have studied the spatial organization of two-dimensional filament networks, confined to the surface of a cylinder. We show that the curvature acts as a field along the axial direction. A simple ring model reveals that this field brings about a bifurcation in the optimal orientation of a polymer that encircles the cylinder: below a critical radius $R^\star$, two degenerate, slanted ground states emerge whereas for radii larger than $R^\star$, only the orientation perpendicular to the cylinder axis is stable. When instead 2D networks of polymers are wrapped around cylinders, this removes the sharp bifurcation structure of the ring model, but its general characteristics persist: ordered, cross-hatched states appear at smaller radii, where the effects of curvature along the substrate are largest. The circumferentially ordered state is prohibited by the geometry and the fixed boundary conditions, and is supplanted by a disordered state in which all angles $\phi$ are equally likely. The latter is understandable, as at infinite radius the effects of curvature drop out. The effects we observe resemble those encountered in various biological settings e.g.\ collagen networks in the growing notochord and in arteries. While certainly not the only physical or biochemical principle in operation here, we hypothesize that the geometric, curvature-induced cross-hatched order we report here may explain, in part, the emergence of similar symmetries in filamentous biomaterials. In addition, the effects we report here may be used to {\em induce} order in synthetic or biomimetic systems: The instructive effects of curvature could provide a novel and noninvasive route towards preparing substrates to predetermined orientational specifications, and we are currently working to extend the framework presented here to arbitrary curved geometries. 

\begin{acknowledgments}
We thank C.V.C. Bouten, T.H. Smit and J. de Vries for valuable discussions. This work was generously supported by the Institute for Complex Molecular Systems (ICMS) at the Eindhoven University of Technology. 
\end{acknowledgments}

\bibliography{ref}

\end{document}